\documentclass{ws-ijmpd}
\usepackage[super,compress]{cite}
\usepackage{graphicx}
\usepackage{epstopdf}
\usepackage{graphicx}
\usepackage{bm}
\usepackage[latin1]{inputenc}
\usepackage{amsmath}
\usepackage{amsfonts}
\usepackage{subfigure}
\usepackage{amssymb}
\usepackage{makeidx}
\usepackage{color}
\usepackage{mathrsfs}
\RequirePackage[colorlinks,citecolor=blue,urlcolor=blue,linkcolor=magenta]{hyperref}
\usepackage[normalem]{ulem}

\begin{document}

\markboth{Binod Chetry, Jibitesh Dutta \and
 Wompherdeiki Khyllep}
{Thermodynamics of  scalar field models with kinetic corrections}

\catchline{}{}{}{}{}
\title{Thermodynamics of  scalar field models with kinetic corrections}

\author{Binod Chetry}
\address{Department of Mathematics, North Eastern Hill University\\ Shillong, Meghalaya 793022, India\\
	\email{binodchetry93@gmail.com}}

\author{Jibitesh Dutta}

\address{Mathematics Division, Department of Basic Sciences and Social Sciences\\ North Eastern Hill University, Shillong, Meghalaya 793022, India\\
	and
	\\
	Visiting Associate, Inter University Centre for Astronomy and Astrophysics\\
	 Pune 411 007, India\\
	jdutta29@gmail.com, jibitesh@nehu.ac.in}
\author{Wompherdeiki Khyllep}

\address{Department of Mathematics, North Eastern Hill University\\  Shillong, Meghalaya 793022, India\\
	and\\
	Department of Mathematics, St. Anthony's College\\ Shillong, Meghalaya 793001, India \\
	sjwomkhyllep@gmail.com}

\maketitle

\begin{history}
\received{Day Month Year}
\revised{Day Month Year}
\end{history}

\begin{abstract}
	In the present work, we compare the thermodynamical viability of two types of non-canonical scalar field models with kinetic corrections: the square kinetic and square root kinetic corrections. In modern cosmology,  the generalised second law of thermodynamics (GSLT) plays an important role in deciding thermodynamical compliance of a model as one cannot consider a model to be viable if it fails to respect GSLT. Hence, for comparing thermodynamical viability, we examine the validity of  GSLT for these two models.  For this purpose,  by employing the Unified first law (UFL),    we calculate the total entropy of these two models in apparent and event horizons. The validity of GSLT is then examined from the  autonomous systems as the original expressions of total entropy are very complicated. Although, at the background level, both models give interesting cosmological dynamics, however, thermodynamically we found that the square kinetic correction is more realistic as compared to the square root kinetic correction. More precisely,  the  GSLT holds for the square kinetic correction  throughout the evolutionary history except only during the radiation epoch where the scalar field may not represent a true description of the matter content. On the other hand, the square root kinetic model fails to satisfy the GSLT in major cosmological eras.
\end{abstract}


%\ccode{PACS numbers:}

\section{Introduction} 
Thermodynamical analysis of the gravity theory is an exciting subject of research in the modern cosmological context.  A fundamental relation between gravity and thermodynamics was derived in the black hole thermodynamics setting \cite{Hawking:1974sw,Bekenstein:1973ur,Jacobson:1995ab,pad,Akbar:2006er,Cai:2005ra,Paranjape:2006ca}. Initially, most of the thermodynamical studies have been focused only on the stationary black holes. Later on, Hayward introduced an approach which deals with the thermodynamical analysis of a dynamical black hole in the Einstein gravity theory \cite{Hayward:1994bu, Hayward:1997jp, Hayward:2004dv,Hayward2}. This approach involves the concept of trapping horizon of a dynamical black hole and also led to an equivalence between Einstein field equations and UFL. However, one must add an extra term called entropy production term for the equivalence to hold in the context of modified gravity theories \cite{Cai1,Akbar:2006kj}.  It is worth mentioning that the UFL is a very rich concept, it can reduce either to the Bondi energy loss equation, or to the first law of relativistic thermodynamics,  or to the first law of thermodynamics for a dynamical black hole depending on the directions where it is projected \cite{Cai1}.

In addition to the first law of thermodynamics, GSLT is one of the fundamental principles of physical thermodynamics. It is worth noting that the  GSLT for a cosmological model should be fulfilled throughout the evolution of the Universe for a model to constitute a perfect thermodynamical system. In the literature, most of the thermodynamical studies for different gravity theories have been performed by using the apparent horizon as a boundary.  Furthermore, while the thermodynamical system on the apparent horizon behaves like a physical Bekenstein system, it is unphysical on the event horizon in the standard gravity theory \cite{Wang:2005pk}. However using various alternative formulations of the Hawking temperature, it was found that Universe bounded by an event horizon forms a perfect thermodynamical system in the Einstein gravity\cite{Saha:2012nz,Chakraborty:2012eu}. 

Recently, there are extensive work on thermodynamical analysis in different gravity theories including $f(R)$-theory \cite{Dutta:2016sgj,Karami:2012hq,MohseniSadjadi:2007zq}, $f(G)$-theory \cite{Sharif:2014esa}, scalar-tensor theory \cite{Karami:2014tsa,Chetry:2018tqr}, Lovelock theory \cite{Cai:2003kt,Akbar:2006kj}, braneworld theories \cite{Sheykhi:2009zza}, general relativity based non-canonical scalar field models  \cite{Das:2015xna, Jawad:2018bpy}.

One of the difficult tasks in analyzing the thermodynamical system is to define the entropy and temperature on the surface of a trapping horizon. In general, these are derived from the black hole thermodynamics in the framework of standard gravity theory, but it is not so in the modified gravity context. There are various forms of temperature used in the literature. One of the well-known form of the horizon temperature is the extended Hawking temperature which plays a vital role in making perfect Bekenstein system in different gravity theories. Note that the extended Hawking temperature is a generalization of the Hayward-Kodoma temperature and the Cai-Kim temperature \cite{Faraoni:2015ula,Moradpour:2015vpa,Cai:2005ra}. In the literature, the validity of  GSLT has been studied by extracting the thermodynamical parameters (extended Hawking temperature and modified entropy) from the UFL in different gravity theories on apparent/event horizon \cite{Dutta:2016sgj,Mitra:2015jha,Saha:2015gha,Mitra:2015wba,Mitra:2015jqa,Mitra:2015nba,Saha:2012nz,Chakraborty:2012eu,Moradpour:2015vpa}. They found that by employing the UFL, the validity of various thermodynamical laws is improved for the Universe bounded by any dynamical horizon irrespective of the choice of the cosmic fluid on the horizon. These interesting results motivate us to study the validity of GSLT in some class of non-canonical scalar field models for the Universe bounded by event/apparent horizon using an extended Hawking temperature and modified entropy derived from the UFL.

The non-canonical scalar field models are widely studied in the literature as the contenders of dark energy (DE). The main advantage of a non-canonical scalar field model is that it can solve the coincidence problem without any fine-tuning issues \cite{Lee:2014bwa}.    In the present work, we shall consider two types of non-canonical scalar fields correspond to the square and square root kinetic corrections of the canonical Lagrangians.  The cosmological evolution of these two models at the background level has been studied earlier for the exponential potential \cite{Tamanini:2014mpa} and later extended to a various class of scalar field potentials \cite{Dutta:2016bbs}. It is important to note that both models lead to interesting phenomenology at early and late times in comparison to the standard canonical case. It is worth noting that even though these non-canonical models exhibit interesting cosmological behavior, it cannot be considered viable if it fails to satisfy fundamental thermodynamical laws. Further, it is known that the thermodynamical non-compliance is one drawback of quintessence models \cite{Duary:2019dfu}. Therefore, we anticipate that the kinetic corrections of the quintessence field might somehow alleviate this shortcoming. Motivated by these facts, in this work, we shall investigate the validity of the GSLT for these two types of kinetic correction scalar field models. For this, we shall use the extended Hawking temperature at the surface of the horizon and modified entropy derived by projecting the UFL along the tangent vector to the horizon.  The main purpose of the present work is to thermodynamically compare the square kinetic  and square root kinetic models, particularly with respect to the validity of GSLT.

The paper is organized as follows: In Sec. \ref{nc}, we present the basic cosmological equations of the kinetic correction non-canonical scalar field model. Next, in Sec. \ref{gslae}, we extract the modified horizon entropy from the UFL. In Sec. \ref{TA}, we discuss the validity of GSLT on the apparent and event horizons for two non-canonical scalar field Lagrangian. Finally, in the last section, we discuss the conclusion of the present work.

In the present work, we shall consider units where  $8\pi=1,~G=1$ and use $(-,+,+,+)$ signature convention of the metric.

\section{Non-canonical scalar field model and basic cosmological  equations}\label{nc}
The general action of a minimally coupled scalar field model can be written as
\begin{equation}\label{eq1}
S=\int d^4x\sqrt{-g}\left[\frac{R}{2}+\mathcal{L}_{\phi}+\mathcal{L}_m\right],
\end{equation}
where $g,~R,~\mathcal{L}_m~ {\rm and}~\mathcal{L}_{\phi}$ denote the determinant of the metric $g_{\mu\nu}$,  the Ricci scalar, the matter Lagrangian and the scalar field Lagrangian respectively. In this work, we specifically focus on the non-canonical scalar field Lagrangian of the form  
\begin{equation}\label{eq2}
\mathcal{L}_{\phi}=V f(B) \,,
\end{equation}
where $V$ stands for the  potential of the scalar field $\phi$ and $f$ is  an arbitrary function of $B$  with
\begin{equation}\label{eq3}
B=\frac{X}{V} \, \quad\text{and}\quad X=-\frac{1}{2}g^{\mu\nu}\partial_{\mu}\phi\partial_{\nu}\phi \,.
\end{equation}    
Clearly, one can recover the canonical case for $f(B)=B-1$. In the literature such scalar field model has been extensively studied as an alternative model of dark energy which is well motivated  from high energy physics \cite{Tsujikawa:2004dp,Piazza:2004df}. From the observational point of view, we shall consider a spatially flat, homogeneous and isotropic FRW universe \cite{Miller,Ade,Ade:2015xua} 
\begin{eqnarray}
ds^2 &=& -dt^2 + a^2(t)dr^2 + R^2 d\Omega_{2}^2\nonumber\\
&=& h_{ab}dx^a dx^b + R^2 d\Omega_{2}^2,  \label{eq5}
\end{eqnarray}
with $R=a(t) r$ denotes the area radius ($a(t)$ is the scale factor with cosmic time $t$), $h_{ab}={\rm diag}(-1,a^2(t))$, $d\Omega_{2}^2=d \theta^2+\sin^2 \theta d\phi^2$ is the   2-space metric  with $a, b=0, 1$  and $x^0=t,x^1=r$.
Varying action (\ref{eq1}) with respect to the above  metric tensor  \eqref{eq5}, one can obtain the Friedmann and acceleration equation respectively as \cite{Tamanini:2014mpa}
\begin{eqnarray}
3H^2&=&\rho_m+\rho_\phi \,,\label{eq6}\\
2\dot{H}+3H^2&=&-(p_m+p_\phi)\,,\label{eq7}
\end{eqnarray}
where $H=\dot{a}/a$ is the Hubble parameter, an overdot denotes the derivative with respect to the cosmic time $t$. Also, $\rho_m$ and $p_m$ are the energy density, pressure of the matter respectively which are related as $p_m=w \rho_m$ with $w$ as the matter equation of state ($-1 \leq w \leq 1$). Further,   $\rho_{\phi}$ and $p_{\phi}$ are the energy density and pressure of the scalar field given by
\begin{eqnarray}
\rho_{\phi}&=&\dot{\phi}^2 \frac{\partial f}{\partial B}-V f \,,\label{eq10}\\
p_{\phi}&=&\mathcal{L}_{\phi}=V f \,.\label{eq11}
\end{eqnarray}
Note that for $f={\rm constant}$, the EoS of a scalar field is $w_{\phi}=\frac{p_\phi}{\rho_\phi}=-1$.  In other words the scalar field model behaves as a cosmological constant for constant $f$.  Now, taking the variation of action (\ref{eq1}) with respect to $\phi$ yields 
\begin{equation}\label{eq4}
\left(\frac{\partial f}{\partial B}+2 B \frac{\partial^2 f}{\partial B^2}\right)\ddot{\phi}+3H\dot{\phi}\frac{\partial f}{\partial B}-\left(f-B\frac{\partial f}{\partial B}+2 B^2 \frac{\partial^2 f}{\partial B^2}\right)\frac{d V}{d \phi}=0 \,.
\end{equation}  
In order to better understand the cosmological dynamics described by Eqs. \eqref{eq6}-\eqref{eq4} in detail, one must specify the function $f$.  In the present work, we consider the power-law kinetic corrections to the canonical scalar field case given by the function \cite{Piazza:2004df}  
\begin{equation}\label{eq17}
f(B) = B-1+\gamma B^n \,,
\end{equation}
and hence the corresponding scalar field Lagrangian can be written as
\begin{equation}\label{lp}
\mathcal{L}_{\phi}=\frac{1}{2}\dot{\phi}^2-V+\gamma V\left(\frac{\dot{\phi}^2}{2V}\right)^n\,,
\end{equation}
 where $\gamma,~n$ are two real parameters which is well defined only for $n>0$.  Here, the parameter $n$ gives correction to the canonical case. The main motivation for considering this Lagrangian is to obtain rich phenomenology compared to the canonical case.  While for $n>1$, one can recover canonical case at the late time but modifies the early dynamics, $n<1$ modifies both the  early and late time dynamics. However, for $n=1$ case, the dynamics resembles that of the canonical scalar field after some field redefinition. Further, if one assumes $n>\frac{1}{2}$ and $\gamma \geq 0$, then for the present kinetic correction model, the scalar field energy density and speed of sound are both positive, therefore physically viable \cite{Tamanini:2014mpa}. Hence, the case of  $n=\frac{1}{2}$ being the limiting case is of particular interest.  For higher-order corrections ($n>2$), the corresponding background cosmological equations are very complicated and the analysis is almost impossible to handle even by numerical techniques. Hence, the choice $n=2$ is the simplest choice which results in a little complicated background dynamics but still sufficiently simple to handle. Thus, from the mathematical as well as a physical point of view  $n=\frac{1}{2}$ and 2 seem to be the natural choice.
		
		The background dynamics of the square kinetic correction ($n=2$) with exponential potential has been first studied by Piazza and  Tsujikawa in Ref. \cite{Piazza:2004df}, but the complete dynamical system analysis for both the kinetic correction models ($n=2, \frac{1}{2}$) has been performed by Tamanini in Ref. \cite{Tamanini:2014mpa}. Subsequently, Dutta et al. \cite{Dutta:2016bbs}  extended this  analysis to various classes of scalar field potentials.  It was indeed found that the kinetic correction models ($n=2, \frac{1}{2}$) are sufficiently simple to handle but also complicated enough in comparison to the canonical case leading to new interesting cosmological dynamics at both the early and late times. In the present work we shall consider the case of $n=2$ and $\frac{1}{2}$ separately for thermodynamical compliance:
		
For the square kinetic correction ($n=2$), the early time dynamics is completely different from the canonical one. For instance, the scalar field kinetic dominated solutions  are replaced by a dark matter or radiation dominated solutions, which is in better agreement with  observations.  Therefore, this model can successfully describe the universe  at both the late times (when dark energy dominates) and  early times (when dark matter dominates) \cite{Tamanini:2014mpa}.

In the square root kinetic case ($n=\frac{1}{2}$), the early time dynamics are similar to the canonical case, although super-stiff solutions ($w_{\rm eff}>1$) always appear. However, at late time this model is found to exhibit a quintom behavior and hence is consistent with the latest astronomical observations favoring the crossing of phantom divide line ($w_{\rm eff}<-1$) \cite{Tamanini:2014mpa}, even though $w_{\rm eff}=-1$ lies inside the two-sigma confidence limit \cite{nov}.

 The energy density  (\ref{eq10}) corresponds to the Lagrangian \eqref{lp} becomes
\begin{eqnarray}
\rho_{\phi}&=&V+\frac{1}{2}\dot{\phi}^2+(2n-1)\gamma V\left(\frac{\dot{\phi}^2}{2V}\right)^n\,. \label{lp1} 
\end{eqnarray}
We note here that the Friedmann equation (\ref{eq6}) can be rewritten as 
\begin{equation}\label{eq24}
\Omega_{\phi}+\Omega_{m}=1 \,,
\end{equation} 
where $\Omega_{\phi}=\frac{\rho_{\phi}}{3H^2}$ and $\Omega_{m}=\frac{\rho_m}{3H^2}$ are the scalar field energy density parameter and energy density parameter of matter respectively.

In order to determine the evolution of the present model, one can convert the above cosmological equations into an autonomous system of differential equations with the help the following dynamical variables introduced in \cite{Tamanini:2014mpa,Dutta:2016bbs}
\begin{equation}\label{eq18}
x_1=\frac{\dot{\phi}}{\sqrt{6} H}\,,~~~~ x_2=\frac{\sqrt{V}}{\sqrt{3}H}\,,~~~~ x_3=-\frac{1}{V}\frac{dV}{d\phi} \,.
\end{equation}
Therefore, using the dimensionless variables (\ref{eq18}), the  cosmological equations \eqref{eq6}-\eqref{eq4} associated with the kinetic correction term \eqref{eq17} can be written as following autonomous system of equations:
\begin{eqnarray}
\frac{dx_1}{dN}&=&\frac{1}{\sqrt{6}\left[1+\gamma n(2n-1){x_1}^{2n-2}{x_2}^{2-2n}\right]}\Big[\lbrace1-(2n^2-3n+1)\gamma x^{2n}_1 x^{-2n}_2\rbrace x^2_2x_3\nonumber \\ & &  -3\sqrt{6}x_1(1+\gamma n {x_1}^{2n-2}{x_2}^{2-2n})\Big]+\frac{3x_1}{2}\Big[(1+w)(1+x^2_1-x^2_2 +\gamma x^{2n}_1x^{2-2n}_2 )\nonumber\\ & & -2 w x^2_1 (1+\gamma n {x_1}^{2n-2}{x_2}^{2-2n})\Big]\,,\label{eq19}\\
\frac{dx_2}{dN}&=&\frac{x_2}{2}\Big[3(1+w)(1+x^2_1-x^2_2+\gamma x^{2n}_1x^{2-2n}_2 )  -6 w x^2_1 (1+\gamma n {x_1}^{2n-2}{x_2}^{2-2n})\nonumber\\ & & -\sqrt{6}x_1x_3\Big]\,, \label{eq20}\\
\frac{dx_3}{dN}&=&-\sqrt{6}\,x_1\,x^2_3\left[\Gamma(x_3)-1\right] \,, \label{eq21}
\end{eqnarray}
where
\begin{equation}
\Gamma= V\ \frac{d^2 V}{d\phi^2} \left( \frac{d V}{d \phi} \right)^{-2} ~~~{\rm and}~~~ N=\ln a \,.
\end{equation}
In Sec. \ref{TA}, we shall use the above autonomous system in order to test the validity of GSLT for two types of kinetic corrections: the square kinetic corrections i.e. $n=2$ and the square root kinetic corrections i.e. $n=\frac{1}{2}$, bounded by the apparent and event horizons separately. By employing the dynamical variables (\ref{eq18}) and the constraint equation \eqref{eq24}, one can rewrite the energy density parameters as
\begin{eqnarray}
\Omega_{\phi}&=& x^2_1+x^2_2+(2n-1)\gamma x^{2n}_1x^{2-2n}_2 \,, \label{eq25}\\
\Omega_{m}&=&1-x^2_1-x^2_2-(2n-1)\gamma x^{2n}_1x^{2-2n}_2 \,.\label{25}
\end{eqnarray}
Further the effective equation of state $w_{\rm eff}$ is given by
\begin{eqnarray}
w_{\rm eff}&=&\frac{p_m+p_\phi}{\rho_m+\rho_\phi}=w+(w+1)(x^2_1-x^2_2+\gamma x^{2n}_1x^{2-2n}_2)\nonumber\\ & & -2 w x^2_1 \Big(1+\gamma n {x_1}^{2n-2}{x_2}^{2-2n}\Big)\,.  \label{eq27}
\end{eqnarray} 
 We note here that as obtained in Ref. \cite{Tamanini:2014mpa} for the square kinetic correction model we shall consider the case where $\gamma>0$ as it is physically viable (i.e. the adiabatic sound speed is positive). However, the square root kinetic correction model is found to be not physically viable in some region of the phase space, if $\gamma x_1<0$. Despite this  drawback, it provides interesting cosmological scenarios such as de-Sitter solution, early time matter dominated solution, scaling solution, phantom dominated solution etc \cite{Tamanini:2014mpa, Dutta:2016bbs}. For instance, they can provide interesting cosmological scenarios such as de-Sitter solution, early time matter dominated solution, quintessence like solution, scaling solution, superstiff and phantom dominated solution \cite{Tamanini:2014mpa, Dutta:2016bbs}. However, some of the cosmic evolutions exhibit  by these models are generically plagued with some perturbational instabilities. Nevertheless, there might be alternative descriptions of nature that produce similar
	background dynamics (see for e.g Ref. \cite{Boehmer:2015sha}), thereby still making the thermodynamical analysis of this paper physically relevant. The interesting  background cosmological features of these non-canonical scalar field models motivate us to analyze the viability of these models from the thermodynamical perspective.  For the thermodynamical analysis, we shall use the extended Hawking temperature and the modified entropy as it  improves the condition for the validity of GSLT.
Therefore, in the next section, we shall derive the modified entropy from the UFL.

	\section{Horizon entropy from the Unified First Law}\label{gslae}
In the framework of universal thermodynamics, Hayward was the first to propose the UFL in order to study the thermodynamical features of a dynamical black hole in the Einstein gravity theory \cite{Hayward:1994bu, Hayward:1997jp, Hayward:2004dv,Hayward2}. He introduced the concept of trapping horizon of a black hole in the Einstein gravity for spherically symmetric spacetimes and subsequently established the equivalence relation between Einstein field equations and the  UFL. It is worth to note that the projection of the UFL along the trapping horizon yields the first law of thermodynamics for the dynamical black hole  \cite{Cai1, Akbar:2006kj, Cai2}.  From the thermodynamical and cosmological point of view, our homogeneous and isotropic FRW Universe can be treated as a non-stationary spherically symmetric spacetime with inner trapping horizon only, coinciding with the apparent horizon.  Hence, it is possible to study the thermodynamical laws associated with FRW based models using UFL.

In terms of the double null coordinates ($\xi ^{\pm}$), the FRW line element \eqref{eq5} can be written as \cite{Cai1}
\begin{equation}
ds^2=-2d\xi^+ d\xi^- +R^2 d\Omega_{2}^2,
\end{equation}
where 
\begin{equation}
\partial _{\pm}=\frac{\partial}{\partial \xi ^{\pm}}=-\sqrt{2}\left(\frac{\partial}{\partial t} \mp \frac{1}{a}\frac{\partial}{\partial r}\right)
\end{equation}
are the future pointing null vectors. 
From the definition of surface gravity ($\kappa_X$), one can obtain its expression for any horizon $X$ having radius $R_{\rm X}$ as
\begin{equation}
\kappa _{\rm X}=-\left(\frac{R_{\rm X}}{R_A}\right)^2\left(\frac{1-\epsilon}{R_{\rm X}}\right)\,.
\end{equation}
In the above expression the suffix $X$ denotes the type of horizon i.e., $X=A$ for the apparent horizon or $X=E$ for the event horizon and $\epsilon =\frac{\dot{R}_A}{2HR_A}$ \cite{Mitra:2015jqa}. Further, the extended Hawking temperature can be written in terms of surface gravity as \cite{Dutta:2016sgj,Mitra:2015jqa,Mitra:2015nba}
\begin{equation}\label{exT}
T_X=\frac{\vert \kappa_X \vert}{2 \pi}\,.
\end{equation}
Now the modified Friedmann Eqs. (\ref{eq6}) and (\ref{eq7}) can be written as
\begin{eqnarray} 
3H^2 &=& \rho_t\,, \label{mfe1}\\
2\dot{H}+3H^2 &=&-p_t\,,\label{mfe2}
\end{eqnarray}
where $\rho_t=\rho_{\rm m} +\rho_{\phi}$ and $p_t=p_{\rm m}+p_{\phi}$ are the total energy density and the thermodynamic pressure. The energy supply vector or energy flux $\psi$ and the work density $W$ are defined as \cite{Hayward:1994bu, Hayward:1997jp, Hayward:2004dv, Hayward2, Cai1, Akbar:2006kj, Cai2}
\begin{equation}
\psi_a=T_a^b\partial_b R+W\partial_a R~,~~~W=-\frac{1}{2}T^{ab}h_{ab}\,,
\end{equation}
where $T_{ab}$ is the energy momentum tensor. Therefore,  for the above modified  Friedmann Eqs. (\ref{mfe1}) and (\ref{mfe2}), the expression of $W$ and $\psi$ become
\begin{eqnarray}
W &=& \frac{1}{2}(\rho_t-p_t)=\frac{1}{2}(\rho_{\rm m} -p_{\rm m}) + \frac{1}{2}(\rho_{\phi}-p_{\phi}) \nonumber \\
&=& W_{\rm m}+W_{\phi}\,,\\
\psi &=& \psi _{\rm m} + \psi _{\phi} \nonumber \\
&=& \left\lbrace -\frac{1}{2}(\rho_{\rm m} +p_{\rm m})HRdt+\frac{1}{2}(\rho_{\rm m} +p_{\rm m})adr \right\rbrace + \Big\lbrace -\frac{1}{2}(\rho_{\phi} +p_{\phi})HRdt \nonumber \\  & & +\frac{1}{2}(\rho_{\phi} +p_{\phi})adr \Big\rbrace .\label{psi}
\end{eqnarray}
Now Eq. \eqref{mfe1} i.e. the ($0$,$0$) component of the modified Einstein's field equation  can be written in the form of the UFL \cite{Hayward:1994bu, Hayward:1997jp, Hayward:2004dv}
\begin{equation} \label{ufl}
dE_X =A_X \psi +WdV_X\,,
\end{equation}
where $E_X$ is the total energy on the horizon of radius $R_X$, whose volume is $V_X$ and surface area is $A_X$.
On projecting the Eq. (\ref{ufl}) along the tangent vector ($\xi_X$) to the surface of the horizon, the  UFL reduces to the first law of thermodynamics at the horizon which takes the form \cite{Cai1, Akbar:2006kj, Cai2}
\begin{equation} \label{uflah}
\langle dE_X, \xi_X \rangle = \kappa_X\langle dA_X, \xi_X \rangle +\langle WdV_X, \xi_X \rangle .
\end{equation}
It is noted that as the heat flow $\delta Q$ is obtained by projecting the pure matter energy supply $A_X \psi _m$ on the  horizon, using Clausius relation (i.e. $\delta Q =T_X dS_X$) we get
\begin{equation} \label{cl}
\delta Q=\langle A_X\psi _m, \xi_X \rangle = \kappa_X\langle dA_X, \xi_X \rangle -\langle A_X \psi _\phi, \xi_X \rangle .
\end{equation}
Now using Eqs. \eqref{exT}, (\ref{psi}) and (\ref{cl}),  $\delta Q$ can be written explicitly as 
\begin{equation}
\delta Q=\langle A_X \psi _m, \xi_X \rangle = T_X \Big\langle \frac{R_X}{4}dR_X-\frac{1}{8} H R_{X}^{4}(\rho _{\phi}+p_\phi)dt, \xi_X \Big\rangle  .
\end{equation}
In ($r$,$t$) coordinates, the tangent vector along the apparent (i.e. $\xi_A$ ) and event horizon (i.e. $\xi_E$)  can be respectively written as
\begin{equation}
\xi_A =\frac{\partial}{\partial t}-(1-2\epsilon)H r \frac{\partial}{\partial r}\qquad \text{and}\qquad \xi_E =\frac{\partial}{\partial t}-\frac{1}{a} \frac{\partial}{\partial r}\,.
\end{equation}
On comparing with the Clausius relation, the  entropy on the apparent/event horizon (obtained by integrating its differential form) can be written as
\begin{eqnarray}
S_A &=& \frac{A_A}{4}-\frac{1}{8}\int HR_{A}^{4}(\rho _\phi+p_\phi)dt\,,\label{mb1} \\
S_E&=&\frac{A_E}{4}-\frac{1}{16} \int \left(\frac{R_{A}^{2}R_E}{1-\epsilon}\right)\left(\frac{HR_E+1}{HR_E-1}\right) (\rho _{\phi}+p_{\phi})dR_E\,, \label{mb2}
\end{eqnarray}
respectively.  It can be seen from Eqs. \eqref{mb1}, \eqref{mb2} that the entropy on both  the  horizons (apparent/event) is nothing but the usual Bekenstein entropy together with a correction term in the integral form. Using these horizon entropies, GSLT has been extensively studied on the apparent and event horizons in different gravity theories \cite{Mitra:2015wba,Saha:2015gha}. It is worth noting that modified horizon entropy obtained from UFL along with the associated extended Hawking temperature  usually leads to a realistic thermodynamical system for various   gravity theories on both the horizons. Thus, in the next section, we shall analyze the thermodynamical behavior of the two non-canonical scalar field models by examining the validity of GSLT using modified horizon entropy and temperature extracted from the UFL.

\section{Thermodynamical analysis}\label{TA}
In this section, we derive the rate of change of the total entropy and then examine the validity of GSLT for the square and square root kinetic corrections. It is well known from the GSLT  that the total entropy ($S_T$) of a physical system must be a non-decreasing function of time i.e. $\dot{S}_T\geq 0$. The total entropy is actually the sum of horizon entropy ($S_h$) and entropy of the fluid inside the horizon ($S_f$).  The entropy of the fluid (i.e. a barotropic matter, $S_f=S_m$) inside the horizon can be extracted from Gibb's equation given by \cite{Wang:2005pk}
\begin{equation}
T_{m}{\rm d}S_{m}={\rm d}E_{X}+p_{m}{\rm d}V_X\,,
\end{equation}
 where $T_{m}$ is the temperature of the fluid. In the present work, we assume that temperature of matter inside the horizon and on the horizon are approximately equal by the local equilibrium hypothesis \cite{Cai2,Izquierdo:2005ku,bcd,db,jbd,Lymperis:2018iuz}.  It is worth mentioning that this hypothesis remains as a conjecture which holds only in a very ideal cosmological setup, as the temperature of the matter content and that of the horizon evolve under different laws almost throughout the expansion history of the Universe. In particular, the hypothesis may not hold during the early radiation era, as the temperature of cosmic microwave background differs from that of the horizon by order of 31  \cite{Mimoso:2016jwg}.  Therefore,  the motivation for considering this hypothesis is to avoid the mathematical complexity associated with the non-equilibrium thermodynamics. However, the following results still imply that the hypothesis cannot be completely discarded:
\begin{itemize}
		\item Thermal equilibrium still might occur at the late times, when the matter component and the horizon will interact for a long time. In Ref. \cite{Mimoso:2016jwg}, it was found that the temperatures of non-relativistic matter and DE might approximately be in equilibrium with that of the horizon. If the equilibrium is reached, then it will stay practically so in the near future. Hence, we have to keep in mind that our results hold in the matter and DE dominated era of the universe  \cite{Lymperis:2018iuz}.
		
		\item The heat flow contribution between the horizon and matter fluid in GSLT  is negligible (of the order $10^{-7}$) and hence the local equilibrium thermodynamics is still preserved \cite{kara}.

		\item  If the temperature of matter and the horizon differs then there is a spontaneous flow of energy between the horizon and matter (vice versa) which in turn may deform the FRW geometry \cite{Izquierdo:2005ku}.
		
		\item  Further as the expression of dark energy temperature is unknown, the assumption of local equilibrium hypothesis is quite economical for analyzing the validity of GSLT.
\end{itemize}
Hence, the local equilibrium hypothesis is not unjustified. By assuming this hypothesis, the rate of change of the matter entropy on the horizon $(\dot{S}_{\rm m})$ is given by
\begin{equation}\label{gibs}
\dot{S}_{\rm m}=\frac{4\pi R^2_{\rm X}}{T_{\rm X}}(\rho_{\rm m}+p_{\rm m})(\dot{R}_{\rm X}-HR_{\rm X}).
\end{equation}
On employing the extended Hawking temperature \eqref{exT} in Eq. (\ref{gibs}) and adding with the time derivative of Eqs. (\ref{mb1})/(\ref{mb2}), we obtain total rate of change of entropy  at the apparent $(\dot{S}_{\rm TA})$/ event $(\dot{S}_{\rm TE})$ horizon respectively as
\begin{eqnarray}
\dot{S}_{\rm TA} &=& \frac{R_{\rm A}\dot{R}_{\rm A}}{4}-\frac{R^3_{\rm A} X}{4}\frac{\partial f}{\partial B}+\frac{R^3_{\rm A}}{4(2-\dot{R}_{\rm A})}(\rho_m+p_m)(\dot{R}_{\rm A}-1)\,,\label{tot}\\
\dot{S}_{\rm TE} &=&\frac{R_{\rm E}\dot{R}_{\rm E}}{4}-\frac{X R^2_{\rm A}R_{\rm E}(\dot{R}_{\rm E}+2)}{4(2-\dot{R}_{\rm A})}\frac{\partial f}{\partial B}-\frac{R_{\rm E}R^2_{\rm A}}{4(2-\dot{R}_{\rm A})}(\rho_m+p_m). \label{tot1} 
\end{eqnarray}
As mentioned earlier for the validity of GSLT, we must have $\dot{S}_{\rm TX}\geq 0$ on any dynamical horizon.  In order to comprehend this, with the help of the governing autonomous system \eqref{eq19}-\eqref{eq21}, in what follows, we shall analyze the validity of GSLT by taking a concrete choice of kinetic correction terms on the apparent and event horizons separately.

	\subsection{GSLT with square kinetic corrections}\label{g1}
This section will be devoted to the study on the  validity of GSLT for the square kinetic corrections scalar field. Therefore, the kinetic correction function $f$ takes the form $f=B-1+\gamma B^2$. Thus, in this case, the scalar field energy density parameter ($\Omega_\phi$), the energy density parameter of matter ($\Omega_m$) and the effective EoS parameter are respectively given by
\begin{eqnarray}
\Omega_{\phi}&=&x^2_1+x^2_2+3\gamma\frac{x^4_1}{x^2_2}\,,\label{eq33}\\ 
\Omega_{m}&=&1-x^2_1-x^2_2-3\gamma\frac{x^4_1}{x^2_2}\,,\label{eq34}\\
w_{\rm eff}&=& w-(w-1)x^2_1-(w+1)x^2_2-\gamma (3w-1)\frac{x^4_1}{x^2_2}\label{eq36} \,.
\end{eqnarray}

\begin{figure}
	\centering
	\subfigure[]{%
		\includegraphics[width=8cm,height=6cm]{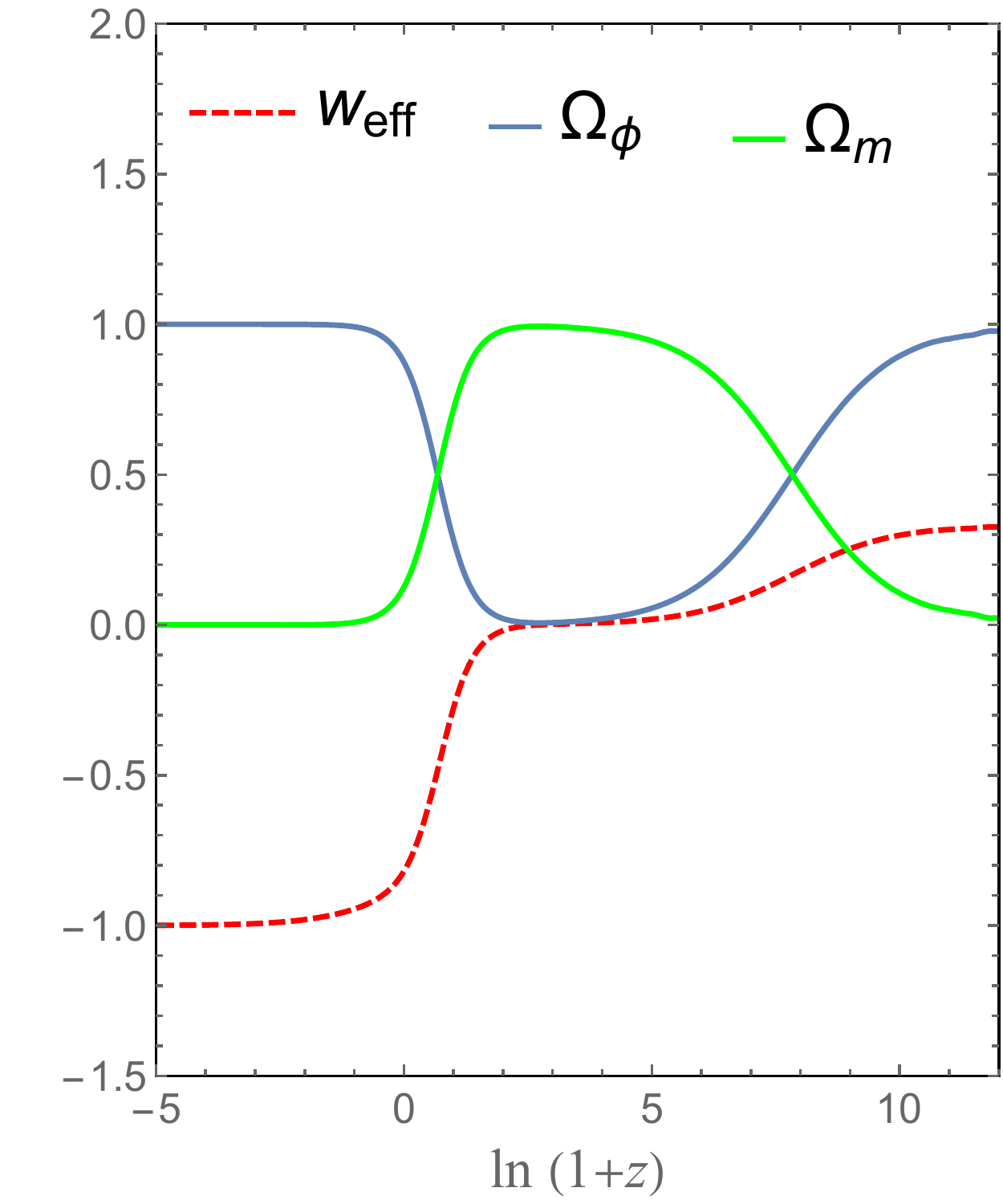}\label{P1}}
	\qquad
	\subfigure[]{%
		\includegraphics[width=8cm,height=6cm]{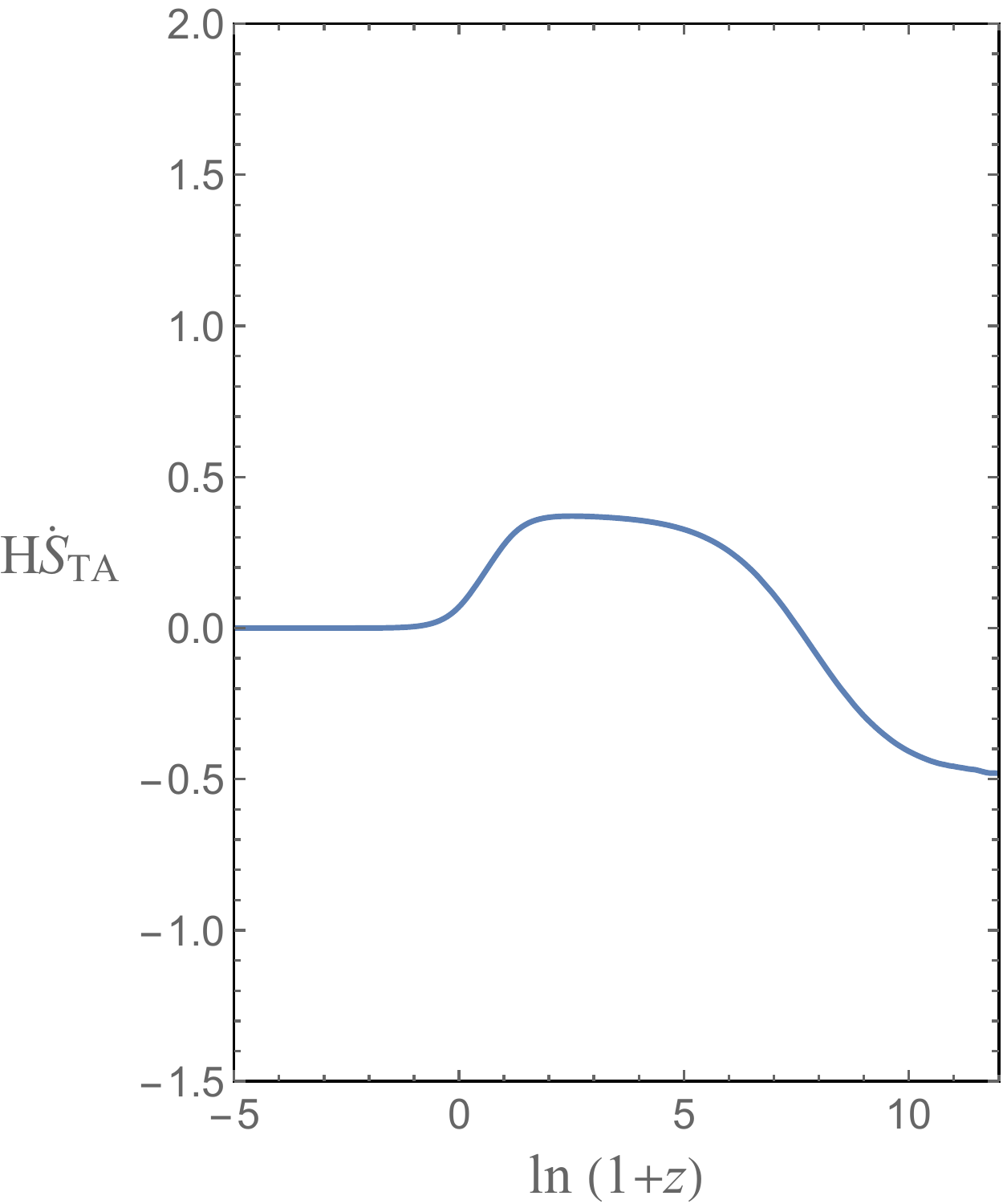}\label{G1}}
	\qquad
	\subfigure[]{%
		\includegraphics[width=8cm,height=6cm]{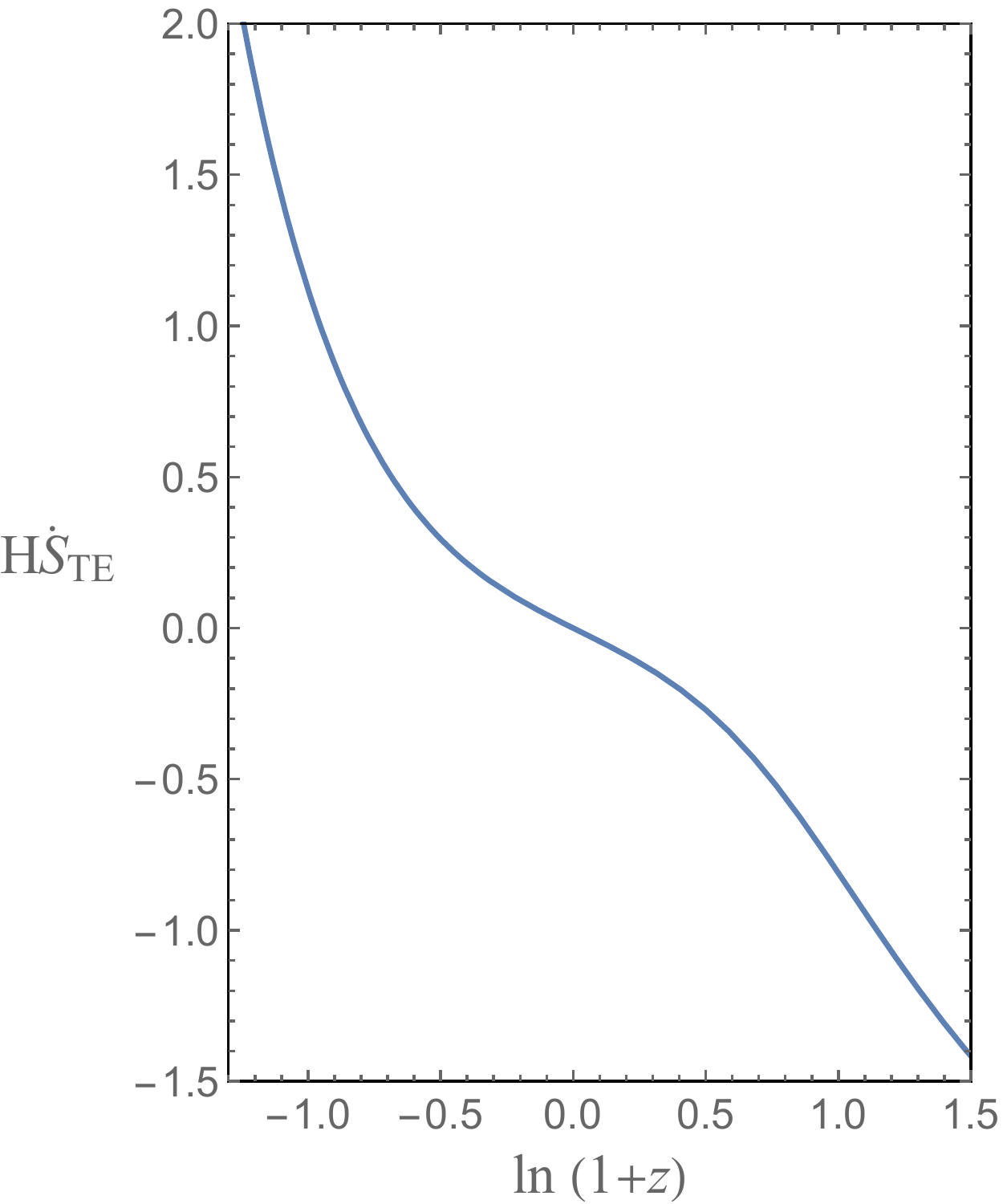}\label{G11}}
	\caption{(a) The evolution of $\Omega_{\phi},~\Omega_{m},~w_{\rm eff}$  versus $z$ for the square kinetic correction model. The evolution of GSLT versus $z$ on the apparent horizon in (b) and event horizon in (c). Here,  we consider a scalar field potential $V=V_0 \sinh^{-\alpha} (\lambda \phi)$ with $w=0,~\gamma=1,~\alpha=-2,~\lambda=0.5$.}\label{fig1}
\end{figure} 
With the help of the governing autonomous system ~\eqref{eq19}-\eqref{eq21}, the evolutionary behavior of $\Omega_{\phi},~\Omega_{m},~ {\rm and}~w_{\rm eff}$ against redshift $z=\frac{a_0}{a}-1$ is given in Fig. \ref{P1}.  Here, $a_0$ denotes the present value of scale factor and it is taken to be unity. It is worth mentioning that for all numerical computations, we have estimated the choice of initial conditions in such a way that the evolution of the Universe is consistent with the present observational data $(\Omega_m=0.3, w_{\rm eff}=-0.7)$ \cite{Ade:2015xua}. It can be seen from Fig. \ref{P1} that the Universe evolves from a scalar field dominated era which effectively scales as radiation ($w_{\rm eff}=\frac{1}{3}$)  and then evolves towards a matter dominated era ($w_{\rm eff}=0$)  before it finally settles as a cosmological constant ($w_{\rm eff}=-1$). Thus the square kinetic correction model describes a true description of the Universe at the background level. It is therefore interesting to discuss the validity of  GSLT  on the apparent and event horizons separately during these significant cosmological eras.\\

\noindent{\bf Case-I: Apparent Horizon}\\
Here, the expression of GSLT (\ref{tot}) for the square kinetic correction model can be expressed in terms of dynamical variables \eqref{eq18} as follows
\begin{eqnarray}
H \dot{S}_{\rm TA} &=& -\frac{3}{8x^2_2}\left[\gamma(3w-1)x^4_1+(w-1)x^2_1x^2_2+(w+1)x^2_2(x^2_2-1)\right]\nonumber \\  & & -\frac{3}{4x^2_2}\left(x^2_1x^2_2+2\gamma x^4_1\right) -\frac{3\Omega_{m}(1+w)}{4}\nonumber \\  & & \times\left[\frac{2x^2_2+3\lbrace\gamma(3w-1)x^4_1+(w-1)x^2_1x^2_2+(w+1)x^2_2(x^2_2-1)\rbrace}{4x^2_2+\lbrace\gamma(3w-1)x^4_1+(w-1)x^2_1x^2_2+(w+1)x^2_2(x^2_2-1)\rbrace}\right]\,.\label{eq40}
\end{eqnarray}
 We note here that the quantity $H \dot{S}_{\rm TA}$ of Eq. \eqref{eq40} is dimensionless as the right hand side is a combination of dimensionless variables. Further, $\Omega_m$ can also be written in terms of dynamical variables \eqref{eq18} by using Eq. \eqref{eq34}. It is worth mentioning that as we consider expanding Universe ($H>0$), the validity of GSLT is equivalent to the condition $H \dot{S}_{TX} \geq 0$ for any horizon $X$. Comparing Fig. \ref{G1} with Fig. \ref{P1}, we see that the GSLT does not hold during the radiation domination but it is satisfied during the matter and  dark energy dominated epoch. Note that at the late time the present model evolves adiabatically as a de-Sitter Universe (i.e. $\dot{S}_{TA}=0$ and $w_{\rm eff}=-1$). Further, we have checked that for a wide range of parameters values $\gamma, \alpha, \lambda$ ($\gamma>0$) the qualitative behavior of the system hardly exhibits any change.  Hence, we present only one plot to determine the validity of GSLT which corresponds with the initial conditions used in Fig. \ref{P1}. Although, in this case, there is a thermodynamic non-compliance during the early radiation era, the square kinetic model still has a scope for a viable model of the Universe. This is because the scalar field may not model correctly the radiation era. Hence,  the present model describes as a good thermodynamical system on the apparent horizon especially during the main cosmological epoch.  \\

\noindent{\bf Case-II: Event Horizon}\\
The radius of the event horizon ($R_{\rm E}$) is given by
\begin{equation} \label{eq41}
R_{\rm E} = a(t)\int^{\infty}_{t}\frac{{\rm d}t'}{a(t')}\,,
\end{equation}
where $a$ and $t$ are the scale factor and the cosmic time
respectively. The above integral converges only if $a(t) \sim t^m$ for $m>1$ i.e. event horizon exists only in the accelerating Universe. From Eq. (\ref{eq41}) we get
\begin{equation}\label{eq42}
\dot{R}_{\rm E}=HR_{\rm E}-1\,.
\end{equation}
In order to determine the validity of GSLT, we shall introduce another dynamical variable $x_4=HR_{\rm E}$ whose corresponding evolution equation is given by
\begin{equation}
\frac{dx_4}{dN}=(x_4-1)+\frac{3x_4}{2x^2_2}[\gamma(3w-1)x^4_1-(w-1)x^2_1x^2_2+ (w+1)x^2_2(x^2_2-1)]\,.\label{eq22}
\end{equation}
Now using dynamical variables \eqref{eq18} together with $x_4$, one can rewrite expression of GSLT  (\ref{tot1}) for the event horizon   as
\begin{eqnarray}
H\dot{S}_{\rm TE}&=&\frac{x_4 (x_4-1)}{4} \nonumber\\ & & -\frac{3}{2}\left[\frac{x_4 (x_4+1)(x^2_1x^2_2+2\gamma x^4_1)}{4x^2_2+3\lbrace\gamma(3w-1)x^4_1-(w-1)x^2_1x^2_2+ (w+1)x^2_2(x^2_2-1)\rbrace}\right] \nonumber\\ & & -\frac{3\Omega_{m}x^2_2x_4(1+w)}{{2[4x^2_2+3\lbrace \gamma(3w-1)x^4_1-(w-1)x^2_1x^2_2+ (w+1)x^2_2(x^2_2-1)\rbrace]}}.\label{eq44}
\end{eqnarray}
Using the autonomous system \eqref{eq19}-\eqref{eq21} together with Eq. (\ref{eq22}), we check the validity of GSLT on the event horizon numerically using Eq. (\ref{eq44}) (see Fig. \ref{G11}). It can be seen by comparing Figs. \ref{P1} and \ref{G11} that the GSLT holds only during the DE domination but fails to satisfy during the early radiation and matter domination. Hence the square kinetic correction model behaves as a perfect thermodynamical system on the event horizon only during the late time accelerated era. Further, as in the previous case, we have checked that the qualitative behavior remains almost  the same  for a various choice of model parameters.

\subsection{GSLT with square root kinetic corrections}
In this section, we examine the validity of GSLT for the square root kinetic corrections of canonical scalar field Lagrangian on  apparent and event horizons separately (i.e. $n=\frac{1}{2}$). Therefore, the corresponding scalar field  Lagrangian (\ref{lp}) become 
\begin{equation}\label{eq45}
\mathcal{L}_{\phi}=\frac{1}{2}\dot{\phi}^2-V+\gamma V\left(\frac{\dot{\phi}^2}{2V}\right)^{\frac{1}{2}}\,.
\end{equation}
In terms of the dynamical variables (\ref{eq18}), the  scalar field energy density parameter, the energy density parameter of matter and the effective EoS parameter are respectively given by
\begin{eqnarray}
\Omega_{\phi}&=&x^2_1+x^2_2\label{eq47}\,,\\
\Omega_{m}&=&1-x^2_1-x^2_2\label{eq48}\,,\\
w_{\rm eff}&=&x^2_1-x^2_2+w(1-x^2_1-x^2_2)+\sqrt{2}\gamma x_1 x_2\label{eq50}\,.
\end{eqnarray} 

\begin{figure}
	\centering
	\subfigure[]{%
		\includegraphics[width=8cm,height=6cm]{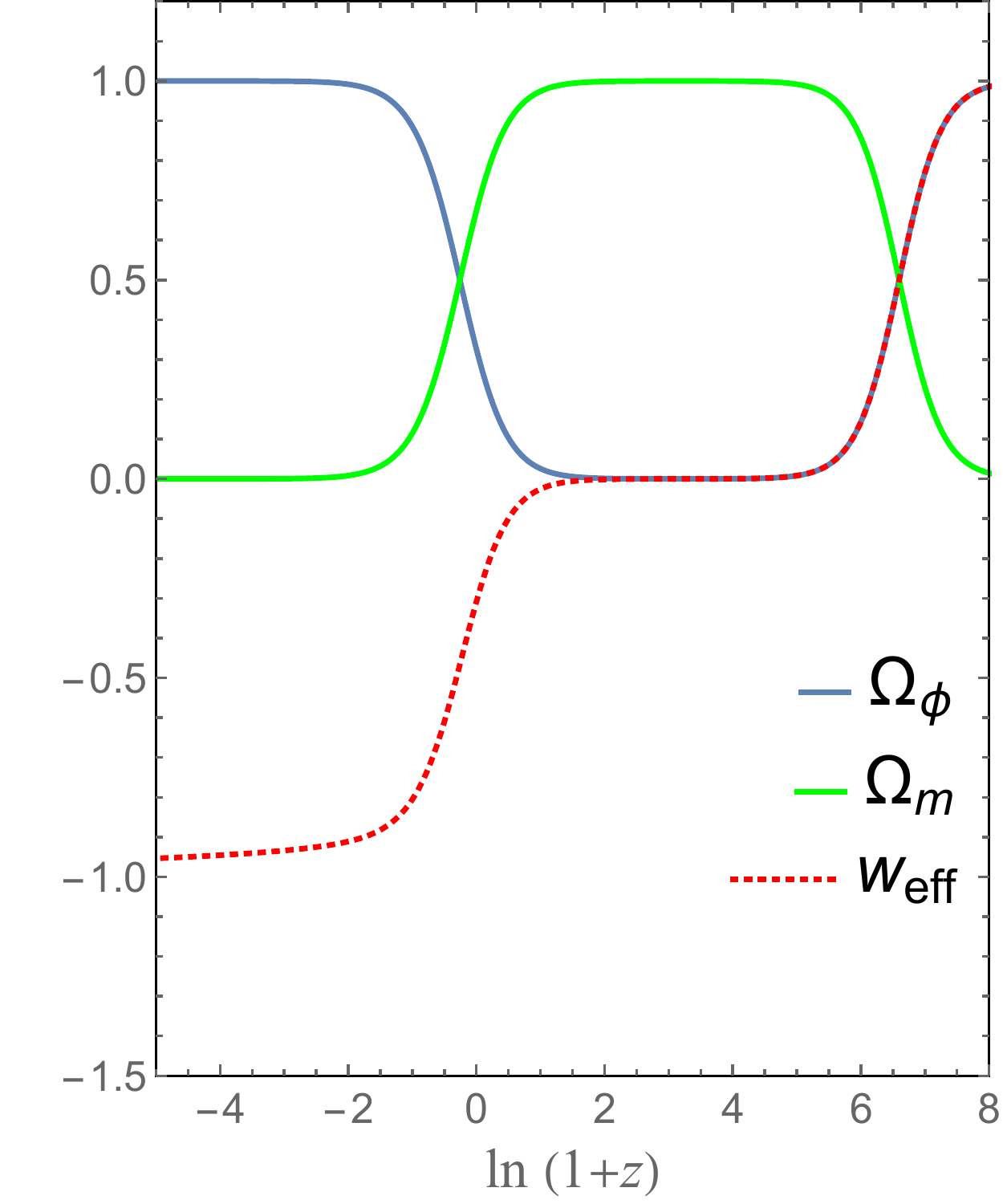}\label{p2}}
	\qquad
	\subfigure[]{%
		\includegraphics[width=8cm,height=6cm]{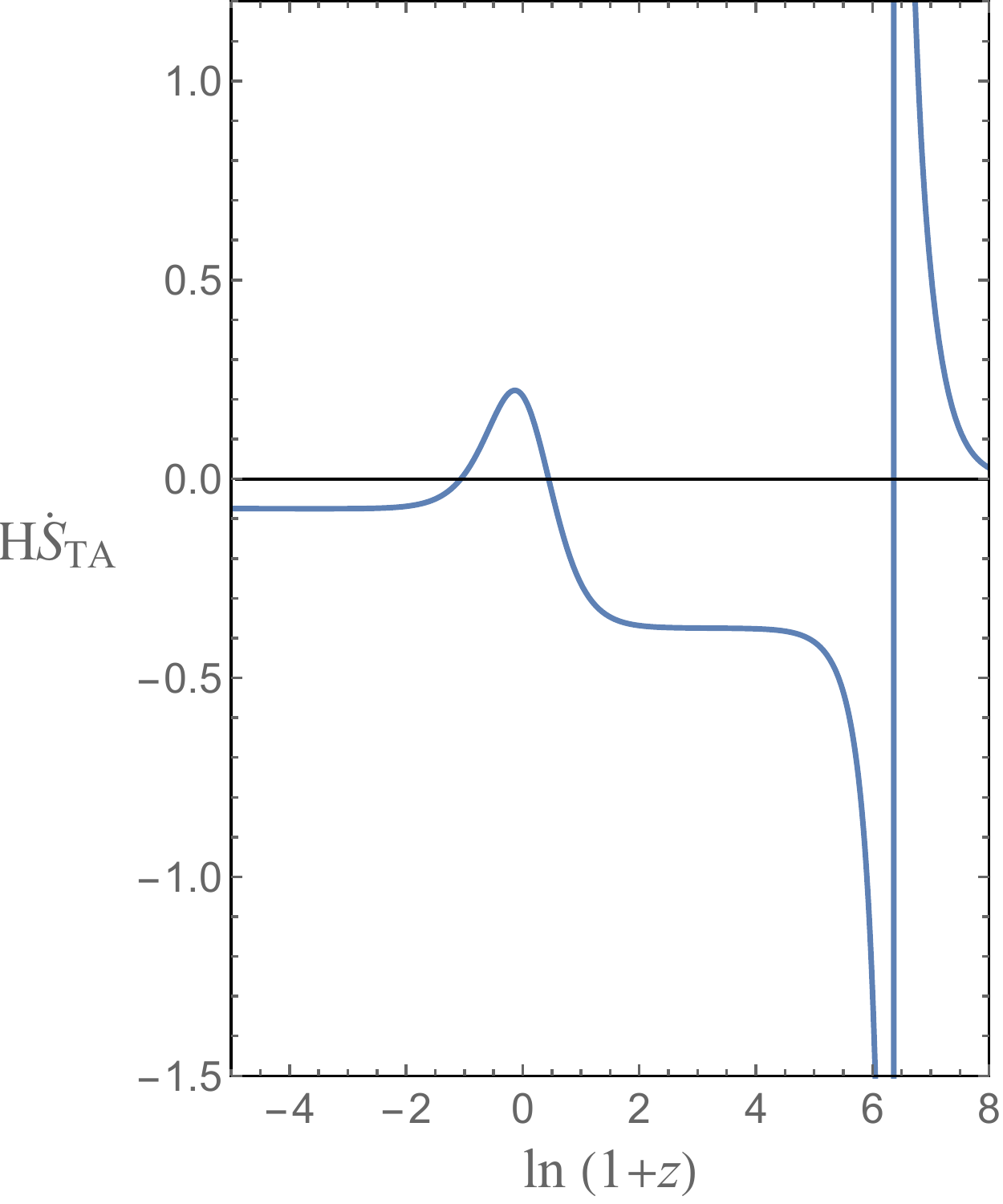}\label{G2}}
	\qquad
	\subfigure[]{%
		\includegraphics[width=8cm,height=6cm]{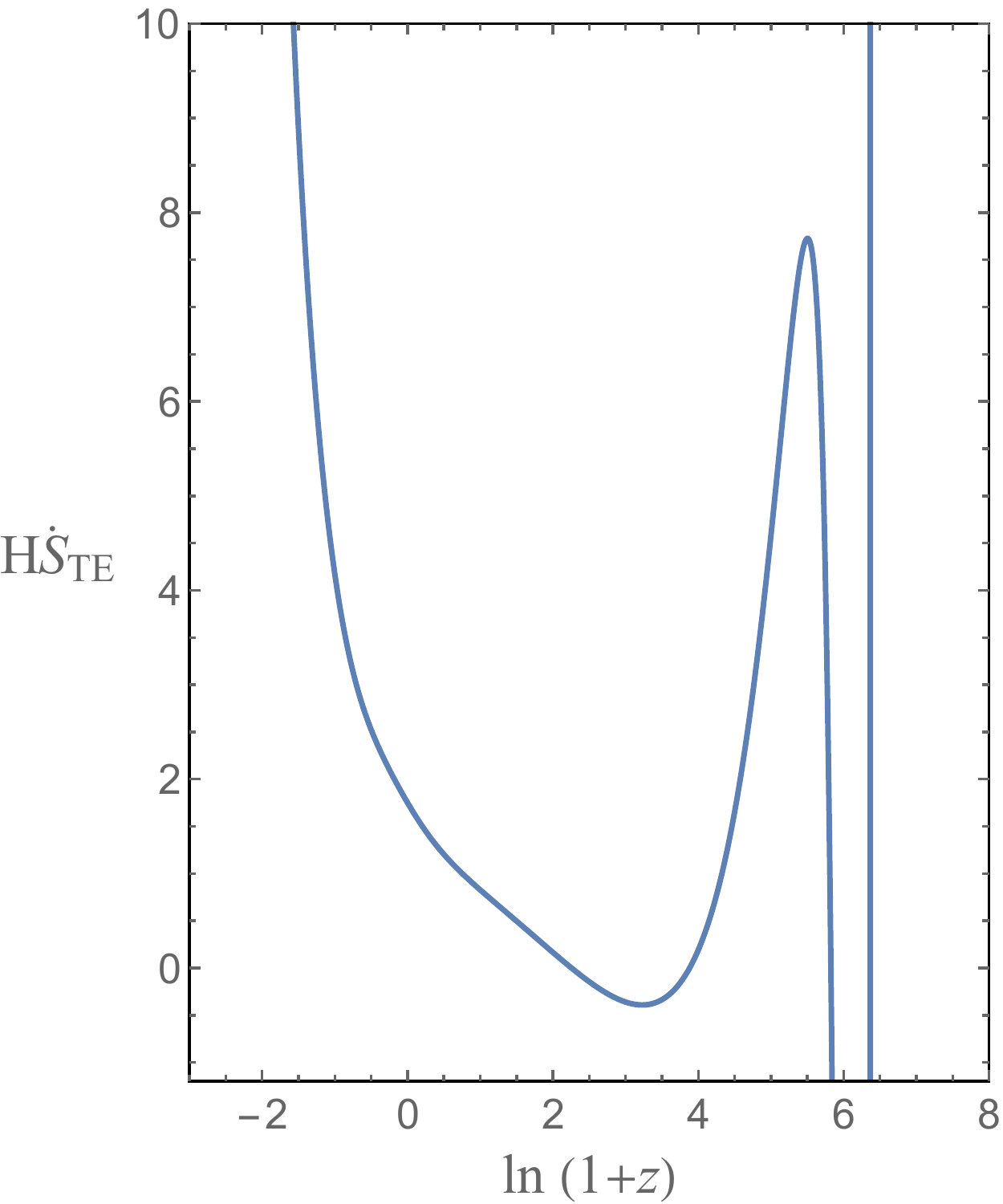}\label{G2_1}}
	\caption{(a) The evolution of $\Omega_{\phi},~\Omega_{m},~w_{\rm eff}$  versus $z$ for the square root kinetic correction model. The evolution of GSLT versus $z$ on the apparent horizon in (b) and event horizon in (c). Here,  we consider a scalar field potential $V=V_0 \sinh^{-\alpha} (\lambda \phi)$ with $w=0,~\gamma=-1,~\alpha=1,~\lambda=0$.}\label{fig2}
\end{figure} 

\begin{figure}
	\centering
	\subfigure[]{%
		\includegraphics[width=8cm,height=6cm]{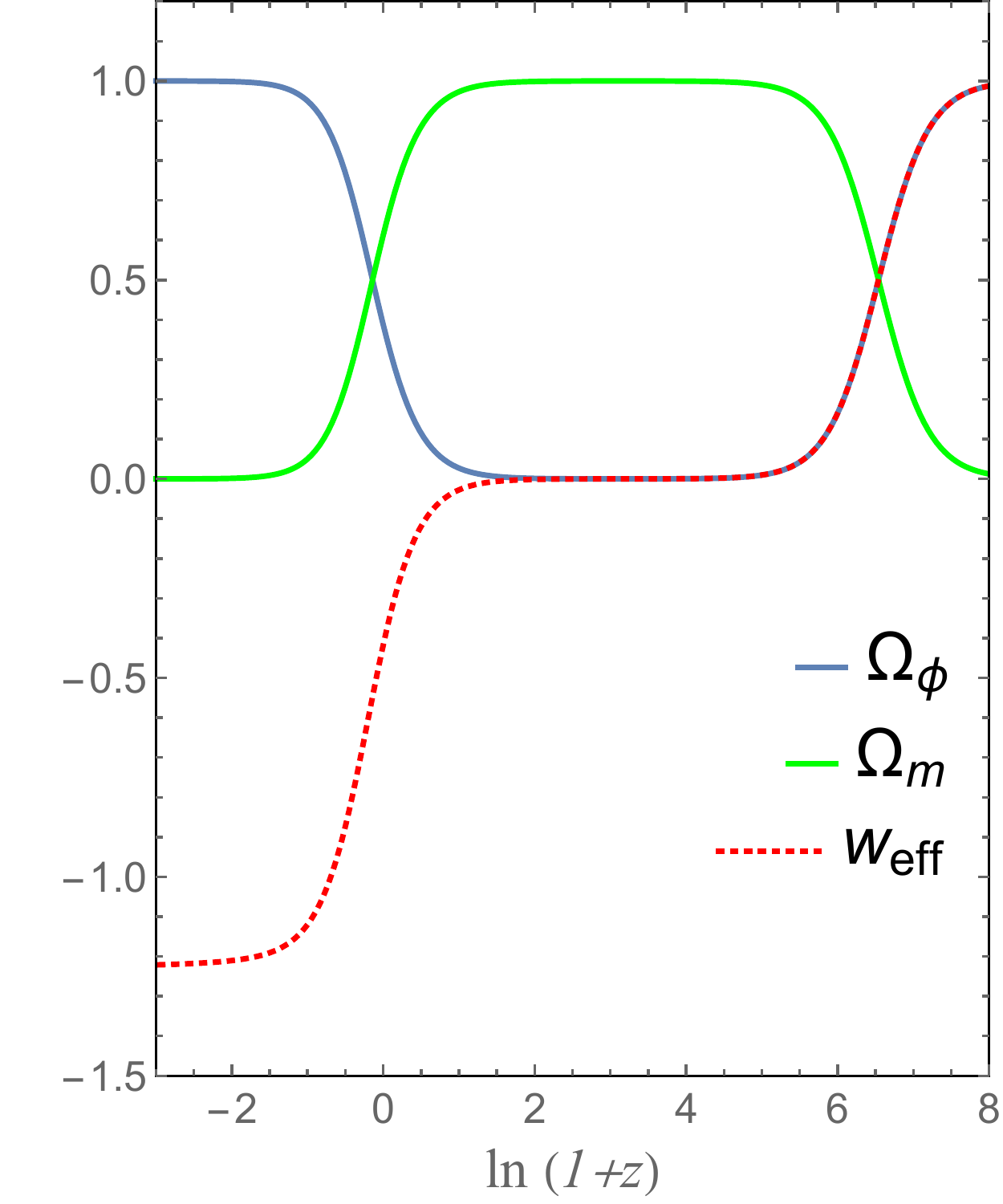}\label{p3}}
	\qquad
	\subfigure[]{%
		\includegraphics[width=8cm,height=6cm]{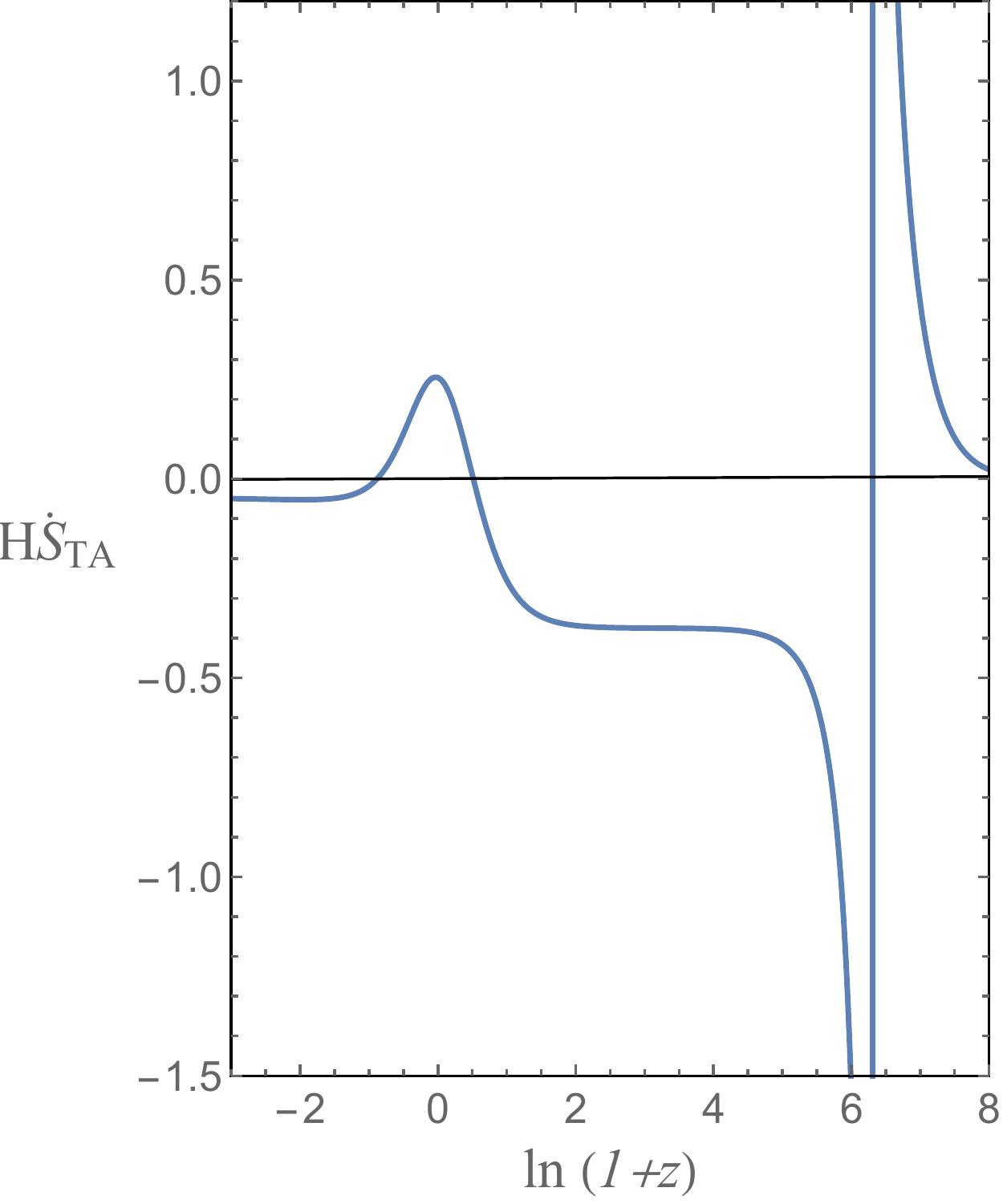}\label{G3}}
	\qquad
	\subfigure[]{%
		\includegraphics[width=8cm,height=6cm]{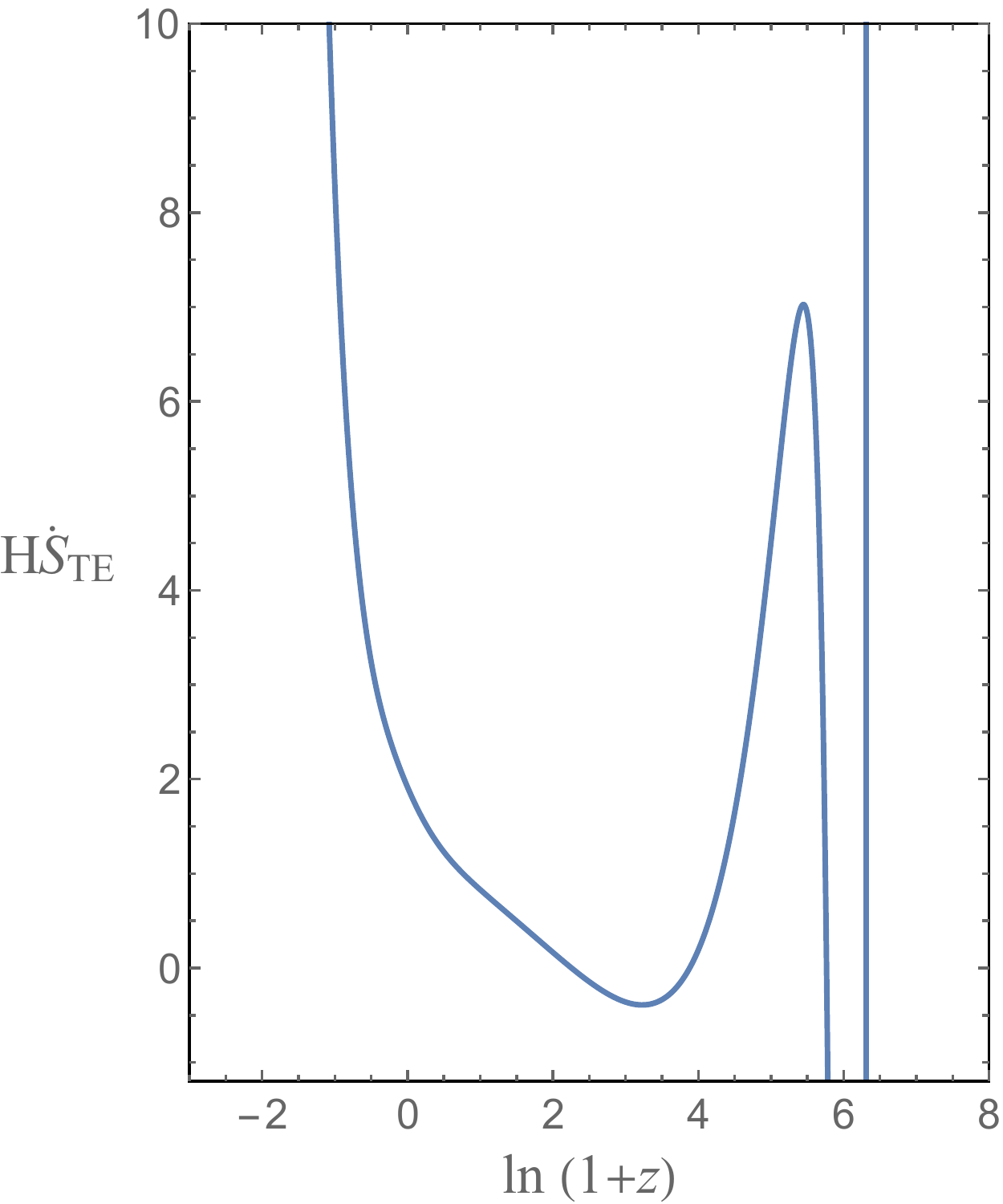}\label{G3_1}}
	\caption{(a) The evolution of $\Omega_{\phi},~\Omega_{m},~w_{\rm eff}$  versus $z$ for the square root kinetic correction model. The evolution of GSLT versus $z$ on the apparent horizon in (b) and event horizon in (c). Here,  we consider a scalar field potential $V=V_0 \sinh^{-\alpha} (\lambda \phi)$ with $w=0,~\gamma=1,~\alpha=-4,~\lambda=\frac{1}{4}$.}\label{fig3}
\end{figure} 
As in the previous section, we can solve the autonomous system of equations  ~\eqref{eq19}-\eqref{eq21} numerically and one can obtain the evolutionary behavior of $\Omega_{\phi},~\Omega_{m}~ {\rm and}~w_{\rm eff}$ for square root kinetic corrections model as shown in Figs. \ref{p2} and \ref{p3}. Depending on the choices of parameter values, it can be seen that the Universe starts to evolve from a stiff matter domination ($w_{\rm eff}=1$) towards a matter domination ($w_{\rm eff}=0$) and then eventually evolves either towards a cosmological constant  ($w_{\rm eff}=-1$) [see Fig. \ref{p2}] or phantom dominated era ($w_{\rm eff}<-1$) [see Fig. \ref{p3}].  As mentioned earlier, even though this model is plagued with some instability problems, however, we proceed ahead with the study of thermodynamical evolution as this model can lead to interesting dynamics at the background level as  in the case of some other physically consistent model \cite{Boehmer:2015sha}. Thus in what follows, we check the validity of GSLT of the square root kinetic correction model on  the apparent and event horizons separately.\\

%\newpage    

\noindent{\bf Case-I: Apparent Horizon}

\noindent	In this case, the expression of GSLT (\ref{tot}) can be written as
\begin{eqnarray}
H\dot{S}_{\rm TA} &=& -\frac{3}{8}\left[(w-1)x^2_1+(w+1)(x^2_2-1)-\sqrt{2}\gamma x_1 x_2\right]-\frac{3}{8}\left(2x^2_1+\gamma x_1x_2\right)\nonumber \\ & &  -\frac{3(1+w)\Omega_{m}}{4}\left[\frac{2+3\lbrace(w-1)x^2_1+(w+1)(x^2_2-1)-\sqrt{2}\gamma x_1 x_2\rbrace}{4+3\lbrace(w-1)x^2_1+(w+1)(x^2_2-1)-\sqrt{2}\gamma x_1 x_2\rbrace}\right]. \label{eq51}
\end{eqnarray}
The validity of GSLT (\ref{eq51}) on the apparent horizon is checked numerically in Figs.  \ref{G2} and \ref{G3}. It can be seen that the GSLT does not hold for the present model in both the matter and DE dominated era. Hence, the present model cannot represent a perfect thermodynamical system on the apparent horizon. Further, it can be seen that at the early times the rate of change of entropy shows some singularity behavior. This pathological behavior associated with the GSLT actually occurs when $w_{\rm eff}=\frac{1}{3}$ which makes the denominator of Eq. \eqref{eq51} to vanish. This is expected as during the radiation era, the scalar field along with pressureless matter may not represent the matter content correctly.\\

\noindent{\bf Case-II: Event Horizon}\\
As in the previous section, we consider another variable $x_4=HR_{\rm E}$ to check the validity of GSLT whose corresponding evolutionary equation is given by
\begin{equation}\label{hr1}
\frac{dx_4}{dN}=(x_4-1)+\frac{3x_4}{2}\left[(w-1)x^2_1+(w+1)(x^2_2-1)-\sqrt{2}\gamma x_1 x_2\right].
\end{equation}
The expression of GSLT (\ref{tot1}) on the event horizon in terms of variables $x_i$'s ($i=1, 2, 3, 4$) can be written as
\begin{eqnarray}\label{eq52}
H\dot{S}_{\rm TE}&=&\frac{x_4(x_4-1)}{4}-\frac{3x_4(x_4+1)(2x^2_1+\gamma x_1x_2)}{4\left[4+3\lbrace(w-1)x^2_1+(w+1)(x^2_2-1)-\sqrt{2}\gamma x_1 x_2\rbrace\right]} \nonumber \\ & & -\frac{3\Omega_{m}x_4(1+w)}{2\left[4+3\lbrace(w-1)x^2_1+(w+1)(x^2_2-1)-\sqrt{2}\gamma x_1 x_2\rbrace\right]}\,.
\end{eqnarray}
From Figs. \ref{G2_1} and \ref{G3_1} one can determine the validity of the GSLT for this model on the event horizon. It can be seen that  GSLT is satisfied only for DE dominated era and also during the earlier and late parts of the dark matter dominated era. Therefore, the present model represents a perfect thermodynamical system during the late time Universe on the event horizon. Hence, from the thermodynamical perspective, the present model is more viable on the event horizon in comparison to the apparent horizon. Again, like the apparent horizon case, here also same pathological behavior associated with the GSLT occurs which corresponds to $w_{\rm eff}=\frac{1}{3}$. We have also checked that similar feature is also obtained for various choices of model parameters. Thus, one can see that the square kinetic correction model is stronger in comparison to the square root correction from the thermodynamical perspective, at least during the matter and late time DE period.

\section{Conclusion}
The non-canonical scalar field models usually provide interesting cosmological dynamics in comparison to the standard canonical one, for e.g. the alleviation of cosmic coincidence problem, the crossing of phantom divide line, etc. Despite, these interesting observational features of the model, one cannot consider a model to be viable if it fails to respect thermodynamical laws. As far as thermodynamical analysis is concerned, modified horizon entropy and extended Hawking temperature usually yield a perfect thermodynamical system in different gravity theories. With this motivation, in this work, we have investigated the validity of GSLT for two non-canonical scalar field FRW models of the Universe bounded by apparent/event horizon separately. 

For the thermodynamical analysis, we have used extended Hawking temperature and extracted modified entropy by projecting the UFL on the direction of the tangential vector to a surface of the horizon. Due to the complicated expression of the GSLT, we have examined its validity by solving the autonomous system \eqref{eq19}-\eqref{eq21}  numerically. For the case of square kinetic corrections, we found that GSLT does not hold during a radiation dominated era on both the horizons. During the matter-dominated era, while the GSLT is satisfied only on the Universe enveloped by the apparent horizon, it fails on the event horizon. However, the GSLT  is satisfied on both the horizons during the DE dominated era (cf. Figs. \ref{fig1}, \ref{fig2}). On the other hand, for the case of square root kinetic corrections, we found that GSLT holds during an early stiff matter dominated era in both the horizons. During the matter-dominated era, the GSLT is not satisfied on both the horizons. However, during the DE dominated era, the GSLT  is satisfied only on the event horizon but not in the case of apparent horizon (cf. Fig. \ref{fig3}).

The square kinetic model can describe the interesting cosmological sequence: radiation $\rightarrow$ matter $\rightarrow$ DE (cf. Fig. \ref{P1}). Although this result is interesting at the background level, it may not be viable from the thermodynamical perspective (cf. Fig. \ref{G1}). However, there is some scope for this model to be viable as the scalar field may not represent a true radiation component. On the other hand, for the case of square root kinetic correction, the model is not thermodynamically compliant in the matter and DE dominated eras in spite of having interesting background dynamics.

In summary, we see that thermodynamically, the square kinetic correction model is more realistic as compared to square root kinetic correction model at least during the matter and late time DE epoch.  It may be noted that our analysis supports the previous results on the cosmological background dynamics of these models. Further, our analysis put further constraints on the viability of the model besides the constraints imposed due to the instability issues presented in Ref. \cite{Tamanini:2014mpa}. We note here that  we have used the local equilibrium hypothesis, however, one can obtain a more vigorous result if non-equilibrium thermodynamics is considered. Further, it is worth mentioning that in this work, we have solved the autonomous system \eqref{eq19}-\eqref{eq21} numerically to examine the thermodynamic compliance. The complete analysis of the autonomous system using dynamical system applications in cosmology \cite{Bahamonde:2017ize} may give a general conclusion (independent of initial conditions) on the thermodynamic viability of a model.  This can be interesting future work. Lastly, a full cosmological perturbation analysis to extract interesting observational signatures against astronomical data are the next logical step to test the viability of these models. We leave it for future work.

\section*{Acknowledgments}
The authors thank the referee for constructive suggestions which lead to the improvement of the work. JD was supported by the Core research grant of SERB, Department of Science and Technology India (File No.CRG/2018/001035) and the Associate program of IUCAA.

\end{document}